\def\beg{\begin{equation}}
\def\eeq{\end{equation}}
\documentstyle[12pt]{article}
\begin{document}
\begin{center}
{\Large{\bf Fractional charge in electron clusters: Mani and
 von Klitzing data of quantum Hall effect- Part II}}
\vskip0.35cm
{\bf Keshav N. Shrivastava}
\vskip0.25cm
{\it School of Physics, University of Hyderabad,\\
Hyderabad  500046, India}
\end{center}

We have calculated the fractional charge of quasiparticles in a 
cluster of electrons. The 61 values have been calculated which 
are exactly the same as the measured values. In a previous eprint
 we have calculated 85 values which are the same as the measured 
values. Thus,
 we have calculated 146 fractional charges which are the same as 
the experimental values. In the case of 61 values, we are able to 
determine the spin of the cluster and hence the number of electrons 
in a cluster. The polarization in a magnetic field is determined. A 
new zero conductivity state is found which is the same  as 
``superresistivity" previously reported in our book. Anderson 
and Brinkman, cond-mat/0302129, have noted that frequencies 
like 5/4$\Omega_c$ do occur. We predict all fractional frequencies 
correctly and report a new state with a zero-frequency mode.

\vskip0.25cm
Corresponding author: keshav@mailaps.org\\
Fax: +91-40-2301 0145.Phone: 2301 0811.
\vskip0.25cm

\noindent {\bf 1.~ Introduction}

     Recently, we[1]  have calculated 85 fractional charges which 
are the same as the experimental values reported by Mani and 
von Klitzing[2]. In all of the 85 values, we have determined the number 
of electrons and their polarization. In some cases, it is clear that 
not all electrons are located at one site. The analysis of the 
data of figures 1-4 of Mani and von Klitzing is given in the previous 
eprint.

     In the present paper, we note the data from figures 5 and 6 of
 Mani and von Klitzing and calculate these 61 values. All of the 61 
calculated
values are the same as the measured values. Such a perfect agreement
 between the calculated and the measured values has never been found 
in the physics problems. The earlier 85 values and the present 61 
values make 146 values for which there is full agreement between the 
calculated and measured values. It was pointed out by Pan et al[3] 
that some of the measured values did not come out from their 
calculations suggesting the inadequacy of their model. Therefore, 
there is a clear need for alternative theoretical possibilities.
 In another eprint, we have shown[4] that the values of Pan et al
 are well predicted by our theory.

\noindent{\bf 2.~~Description}

     From the central part of figure 5 of Mani and von Klitzing, 
we read the following experimental values,
\beg
6/17, 11/31, 9/25, 4/11, 7/19, 10/27, 11/29, 8/21, 5/13
\eeq
\beg
2/5, 7/17, 8/19, 3/7, 4/9, 5/11
\eeq
\beg
16/39, 12/29, 13/31, 10/23, 11/25.
\eeq
The right-hand-side inset gives the following experimental values,
\beg
16/37, 29/67, 23/53, 17/39, 24/55, 25/57, 18/41, 11/25, 4/9
\eeq
and the left-hand-side of figure 5 has the following,
\beg
11/27, 20/49, 16/39, 12/29, 17/41, 18/43, 8/19, 3/7.
\eeq
{\it {\bf Table 1:}The interpretation of fractional charges in 
terms of ${\it l}$ and $s$. The calculated values are the same as
 the experimental values.}
\vskip0.25cm
\begin{center}
\begin{tabular}{ccccc}
\hline
S.No.& ${\it l}$ &  $s$  & $\nu_-$ & $n_e$\\
\hline
1    &   8       & -5/2  &  6/17   &  5\\
2    &  15       & -9/2  & 11/31   &  9\\
3    &  12       & -7/2  &  9/25   &  7\\
4    &   5       & -3/2  &  4/11   &  3\\
5    &   9       & -5/2  &  7/19   &  5\\
6    &  13       & -7/2  & 10/27   &  7\\
7    &  14       & -7/2  & 11/29   &  7\\
8    &  10       & -5/2  &  8/21   &  5\\
9    &   6       & -3/2  &  5/13   &  3\\
10   &   2       & -1/2  &  2/5    &  1\\
11   &   8       & -3/2  &  7/17   &  3\\
12   &   9       & -3/2  &  8/19   &  3\\
13   &   3       & -1/2  &  3/7    &  1\\
14   &   4       & -1/2  &  4/9    &  1\\
15   &   5       & -1/2  &  5/11   &  1\\
16   &  19       & -7/2  & 16/39   &  7\\
17   &  14       & -5/2  & 12/29   &  5\\
18   &  13       & -5/2  & 13/31   &  5\\
19   &  11       & -3/2  & 10/23   &  3\\
20   &  12       & -3/2  & 11/25   &  3\\
\hline
\end{tabular}
\end{center}
\vskip0.25cm

While writing the book[5] we have found that the fractional 
effective charge is given by,
\beg
e_{eff}/e = {{\it l}+{1 \over 2}\pm s\over 2{\it l}+1}
\eeq
When we substitute the values of ${\it l}$ and $s$, we obtain the 
calculated values of the fractional charge. Table 1 shows 20 
calculated values for various values of ${\it l}$ and $s$. All of 
these 20 calculated values are the same as the measured values. Next,
 we calculate 9 values as given in Table 2. These calculated values 
are the same as the experimental values of eq.(4). We calculate 
another set of 8 values given in Table 3. These calculated values are 
the same as the experimental values of eq.(5). All the calculated 
values given 
in Tables 1-3 are thus the same as the experimental values. All of 
these use negative sign for the spin. Hence, these are polarized in 
the magnetic field with all particles parallel to the field with no
 particle with opposite spin. From the spin, we can know the number 
of electrons. Thus there are electron clusters with number of electrons 
1, 3, 5, 7 or 9, in a cluster. All these electron numbers per cluster 
are odd numbers. For even number of electrons, the denominator becomes 
even. For odd number of electrons, the denominator is odd.

{\it {\bf Table 2}: Some of the fractional charges in terms of
 ${\it l}$ and $s$. The calculated values are the same as the 
experimental values.}
\vskip0.25cm
\begin{center}
\begin{tabular}{ccccc}
\hline
S.No.& ${\it l}$ & $s$  & $\nu_-$ & $n_e$\\
\hline
1    &  18       & -5/2 & 16/37   &  5\\
2    &  33       & -9/2 & 29/67   &  9\\
3    &  26       & -7/2 & 23/53   &  7\\
4    &  19       & -5/2 & 17/39   &  5\\
5    &  27       & -7/2 & 24/55   &  7\\
6    &  28       & -7/2 & 25/57   &  7\\
7    &  20       & -5/2 & 18/41   &  5\\
8    &  12       & -3/2 & 11/25   &  3\\
9    &   4       & -1/2 &  4/9    &  1\\
\hline
\end{tabular}
\end{center}
\vskip0.25cm
{\it {\bf Table3}: Some more fractional charges in which the 
calculated values are the same as the experimental values.}
\vskip0.25cm
\begin{center}
\begin{tabular}{ccccc}
\hline
S.No. & ${\it l}$ & $s$  & $\nu_-$ & $n_e$\\
\hline
1     &  13       & -5/2 & 11/27  & 5\\
2     &  24       & -9/2 & 20/49  & 9\\
3     &  19       & -7/2 & 16/39  & 7\\
4     &  14       & -5/2 & 12/29  & 5\\
5     &  20       & -7/2 & 17/41  & 7\\
6     &  21       & -7/2 & 18/43  & 7\\
7     &   9       & -3/2 &  8/19  & 3\\
8     &   3       & -1/2 &  3/7   & 1\\
\hline
\end{tabular}
\end{center}
\vskip0.25cm

Now we take the experimental values from figure 6 of Mani and
 von Klitzing. These measured values are given below:
\beg
1/3, 2/5, 3/7, 4/9, 4/7, 3/5, 2/3, 1, 2, 3, 4,
\eeq
\beg
9/25, 4/11, 7/19, 8/21, 5/13, 7/17, 8/19, 11/25
\eeq
\beg
11/19, 8/13, 7/11, 5/7, 8/11, 7/9, 4/5, 9/11,
\eeq
\beg
9/7, 4/3, 7/5, 8/5, 5/3, 7/3, 8/3, 11/3.
\eeq
The calculation of these values is given  in Tables 4-7.

\vskip0.25cm
{\it {\bf Table 4}: For $s$=$\pm$1/2 both the spin polarizations 
occur. The calculated values of the fractional charges in terms 
of ${\it l}$ and $s$  are the same as those measured. In the first
 line $\nu_-$=0 indicates zero-frequency mode which is characteristic 
of charge-density waves.}
\vskip0.25cm
\begin{center}
\begin{tabular}{ccccc}
\hline
S.No. & ${\it l}$  & $s$ & $\nu_+$ & ${\nu_-}$\\
\hline
1     &  0         &  1/2&     1   &     0\\
2     &  1         &  1/2&     2/3 &     1/3\\
3     &  2         &  1/2&     3/5 &     2/5\\
4     &  3         &  1/2&     4/7 &     3/7\\
5     &  4         &  1/2&     5/9 &     4/9\\
6     & $\infty$   &  1/2&     1/2 &     1/2\\ 
\hline
\end{tabular}
\end{center}
\vskip0.25cm
{\it {\bf Table 5}: Several fractional charges calculated for
 ${\it l}$ and $s$. The calculated values are the same as the 
measured values.}
\vskip0.25cm
\begin{center}
\begin{tabular}{cccc}
\hline
S. No. & ${\it l}$ & $s$  & $\nu_-$\\
\hline
1      & 12        & -7/2 & 9/25\\
2      &  5        & -3/2 & 4/11\\
3      &  9        & -5/2 & 7/19\\
4      & 10        & -5/2 & 8/21\\
5      &  6        & -3/2 & 5/13\\
6      &  8        & -3/2 & 7/17\\
7      &  9        & -3/2 & 8/19\\
8      & 12        & -3/2 & 11/25\\
\hline
\end{tabular}
\end{center}
\vskip0.25cm

Table 4 shows that as long as the spin is $\pm 1/2$, both spin 
polarizations occur except for the small difference due to Boltzmann
 factor. Table 5 shows that only negative spin occurs due to 
polarization in the magnetic field. Similarly, Tables 6 and 7 show
 only the positive spin. In some places large values of spin occur so 
that electrons must be distributed over several sites so we get the 
idea of sites and clusters.

We learn from this exercise that there are electron clusters with a 
small number of electrons per cluster. There are spin polarizations
 in the field so that in some region all spins are aligned in one 
direction only. We determine the number of electrons per cluster.
 [In ref.1 the fraction 29/49 was left out. It is well predicted by 
${\it l}$=24 and $s$=9/2 with positive sign for the spin. Therefore 
85 charges are predicted in this reference].

\noindent{\bf3.~~ Zero conductance, zero charge or superresistivity}.

     When the charge of the quasiparticles is zero, the resistivity
 becomes infinite and hence the conductivity at this point is zero. 
This is the phenomenon of ``superresistivity" as it was called in the
 book[5]. The quantized resistivity is given by,
\beg
\rho_{xy}={h\over ie^2}
\eeq
When $i$=0, $\rho_{xy}$=$\infty$. This gives the divergence in the
 resistivity so that there is ``superresistivity". In an approximate
 representation,
\beg
\sigma_{xx}= \rho_{xx}/\rho_{xy}^2.
\eeq
\vskip0.25cm
{\it {\bf Table 6}: More values of fractional charges in terms of
 ${\it l}$ and $s$. The calculated values are the same as the
 experimental values.}
\vskip0.25cm
\begin{center}
\begin{tabular}{cccc}
\hline
S.No. & ${\it l}$ & $s$ & $\nu_+$\\
\hline
1     &  9        & 3/2 & 11/19\\
2     &  6        & 1/2 &  8/13\\
3     &  5        & 3/2 &  7/11\\
4     &  3        & 3/2 &  5/7\\
5     &  5        & 5/2 &  8/11\\
6     &  4        & 5/2 &  7/9\\
7     &  2        & 3/2 &  4/5\\
8     &  5        & 7/2 &  9/11\\
\hline
\end{tabular}
\end{center}
\vskip0.25cm
{\it {\bf Table 7}: More fractional charges. There are 61 fractional
 values in this paper. All of the calculated values are the same as
 the experimental values.}
\vskip0.25cm
\begin{center}
\begin{tabular}{cccc}
\hline
S.No. & ${\it l}$& $s$  & $\nu_+$\\
\hline
1     &  3       &  7/2 & 9/7\\
2     &  1       &  5/2 & 4/3\\
3     &  2       &  9/2 & 7/5\\
4     &  2       & 11/2 & 8/5\\
5     &  1       &  7/2 & 5/3\\
6     &  1       & 11/2 & 7/3\\
7     &  1       & 13/2 & 8/3\\
8     &  1       & 19/2 & 11/3\\
\hline
\end{tabular}
\end{center}
\vskip0.25cm

When $\rho_{xy}$=$\infty$, $\sigma_{xx}$=0. These are the ``zero
 conductivity" points. The effective charge of the quasiparticles is,
\beg
i= {{\it l}+{1\over 2}\pm s\over 2{\it l}+1}
\eeq
which is zero for ${\it l}+{1\over 2}\pm s$=0. For ${\it l}$=0,
 $s$=1/2 for the negative sign, the charge becomes zero. This is 
because the charge acquires a vector nature due to ${\it l}$ and $s$. 
The effective charge is zero for ${\it l}$=0, $s$=-1/2; ${\it l}$ =1,
 $s$= -3/2;
${\it l}$=2, $s$=-5/2; ${\it l}$=3, $s$=-7/2, ..., etc. These are the
 points of infinite $\rho_{xy}$ and zero conductivity.

     Yang et al[6] have found the points of zero conductivity. In 
eq.(12) making $\rho_{xy}$=$\infty$, $\sigma_{xx}$=0 is consistent 
with the experimental observation of zero conductivity but now 
$\rho_{xx}$
is not known. Let us search for points where $\rho_{xx}$ may also be
 zero. If we use semiquantized values, then by using classical Hall
 effect, we can write $\sigma_{xx}$=$\rho_{xx}(nec/B)^2$ which can give
$\rho_{xx}$=0 points as well as $\sigma_{xx}$=0. Therefore, in the 
semiclassical model both the conductivity as well as the resistivity 
can show  zeros. Recently, several authors have shown interest in this 
problem[7-10]. Anderson and Brinkman[7] have noted  that the frequencies
 of 3/4$\Omega_c$, 7/4$\Omega_c$, 5/4$\Omega_c$ do occur. Our formula
 (6) works very well and predicts all fractions correctly.

\noindent{\bf4.~~ Conclusions}

     There are single electrons which give the fractional charge as
 given in Table 4. There are clusters of 2, 3, 4, ..., 9 electrons. 
The electrons are not all on one site. They are spin polarized. The 
vector nature of charge is exhibited when it is described in terms 
of ${\it l}$ and $s$. There is a state of zero charge and
 ``superresistivity". Along with ref.1, we have calculated 146 
fractional charges in which all of the calculated values are the
 same as the experimentally measured values. The fractional charge 
1/3 occurs in Table 4,in a natural way and when the field is varied, 
its Kramers conjugate with charge 2/3 is found. It is shown 
elsewhere[5] that the calculated temperature dependent spin 
polarization agrees with the experimental data. Similarly, there 
is a Bose-Einstein condensation at half filled Landau level.

    Are there any charge-density waves? Yes, we do have a solution
 with zero frequency, ${\it l}$=0, $s$=-1/2. Are there any
 superconducting states? Yes, clusters of three electrons with all 
spin parallel do not have a Meissner effect but can superflow. At 
very large values of ${\it l}$ it is possible to have singlet pairs.
 For ${\it l}$=0, $s$=1, there
are two solutions, one has a charge of 3/2 and the other -1/2, 
the difference between the two being 2$e$, can exhibits triplet 
type superconductivity but there are clusters so that uniform 
medium is 
not available.

\noindent{\bf5.~~References}
\begin{enumerate}
\item K. N. Shrivastava, cond-mat/0303309.
\item R. G. Mani and K. von Klitzing, Z. Phys. B{\bf 100}, 635(1996).
\item W. Pan, H. L. Stormer, D. C. Tsui, L. N. Pfeiffer, K. W. 
Baldwin and K. W. West, Phys. Rev. Lett. {\bf 90}, 016801 (2003).
\item K. N. Shrivastava, cond-mat/0302610.
\item K.N. Shrivastava, Introduction to quantum Hall effect,\\ 
      Nova Science Pub. Inc., N. Y. (2002).
\item C. L. Yang, M. A. Zudov, T. A. Knuuttila, R. R. Du, L. N. 
Pfeiffer and K. W. West, cond-mat/0303472
\item P. W. Anderson and W. F. Brinkman, cond-mat/0302129.
\item R. G. Mani, J. H. Smet, K. von Klitzing, V. Narayanmurti,
 W. B. Johnson and V. Umansky,  cond-mat/0303034.
\item F. S. Bergeret, B. Huckestein and A. F. Volkov,
 cond-mat/0303530.
\item A. F. Volkov, cond-mat/0302615.
\end{enumerate}
\vskip0.1cm

Note: Ref.5 is available from:\\
 Nova Science Publishers, Inc.,\\
400 Oser Avenue, Suite 1600,\\
 Hauppauge, N. Y.. 11788-3619,\\
Tel.(631)-231-7269, Fax: (631)-231-8175,\\
 ISBN 1-59033-419-1 US$\$69$.\\
E-mail: novascience@Earthlink.net

\end{document}